\documentclass[review]{elsarticle}

\usepackage{lineno,hyperref}

\usepackage[margin=1in]{geometry}

\pdfoutput=1

\usepackage{graphicx}
\usepackage{bm}
\usepackage{dcolumn}
\usepackage{setspace}
\usepackage{amsmath}
\usepackage{fancyhdr}

\usepackage{caption}
\usepackage{subcaption}

\usepackage{color,soul}

\modulolinenumbers[5]

\makeatletter
\AtBeginDocument{\let\hl\@firstofone}
\makeatother

\begin{document}

\begin{frontmatter}

\title{Atomistic mechanisms of twin-twin interactions in Cu nanopillars}




\author[MDTD-IGC]{G. Sainath\corref{mycorrespondingauthor}}
\cortext[mycorrespondingauthor]{Corresponding author}
\ead{sg@igcar.gov.in or mohansainathan@gmail.com}

\author[MDTD-IGC,HBNI]{Sunil Goyal}

\author[MDTD-IGC,HBNI]{A. Nagesha}

\address[MDTD-IGC]{Materials Development and Technology Division, Metallurgy and Materials Group, Indira Gandhi Centre for Atomic Research, Kalpakkam, Tamilnadu-603102, India}

\address[HBNI]{Homi Bhabha National Institute, Indira Gandhi Centre for Atomic Research, Kalpakkam, Tamilnadu-603102, India}

\begin{abstract}

Twinning is an important mode of plastic deformation in metallic nanopillars. When twinning occurs on multiple systems, it 
is possible that twins belonging to different twin systems interact and forms a complex twin-twin junctions. Revealing 
the atomistic mechanisms of how twin-twin interactions lead to different twin junctions is crucial for our understanding of 
mechanical behaviour of materials. In this paper, we report the atomistic mechanisms responsible for the formation of two 
different twin-twin interactions/junctions in Cu nanopillars using atomistic simulations. One junction contains two twin 
boundaries along with one $\Sigma$9 boundary, while the other contains five twin boundaries (five-fold twin). These 
junctions were observed during the tensile deformation of [100] and $[1\bar1 0]$ Cu nanopillars, respectively.\\
  
\end{abstract}


\end{frontmatter}

{\centering
\section*{Graphical Abstract}

 \begin{figure}[h]
 \centering
 \includegraphics[width=10cm]{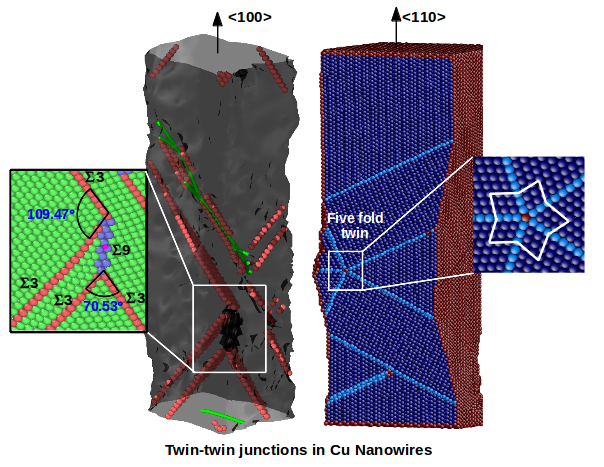}
 \label{abstract}
 \end{figure}
 }


\section{Introduction}

Many experimental and atomistic simulation studies have shown that twinning is an important mode of plastic deformation 
in FCC nanopillars \cite{Cai-Rev,Lee-Nature,Mao-EML} and nanocrystalline materials \cite{New-twinning,Nano-Pt}. Generally, 
twinning occurs in low stacking fault energy (SFE) materials 
such as Au and Cu under conditions that lead to high stresses, such as high strain rates, low temperatures and small size 
\cite{NanoLett,Kiener}. Low SFE increases the separation between leading and trailing partials, which increases the barrier 
for full slip leading to the occurrence of twinning. In nanopillars, the occurrence of twinning has been explained in terms 
of different Schmid factors for leading and trailing partials \cite{Cai-Rev}. Twinning occurs when the Schmid factor for 
leading partial is higher than the trailing partial \cite{Cai-Rev}. This condition is readily met in $<$100$>$ nanopillar 
deforming under compression, $<$110$>$ and $<$111$>$ nanopillars under tension. As a result, numerous studies have reported 
twinning in these orientations and loading conditions \cite{Lee-Nature,Cao-Acta-Cu,Zheng-Nature,Rohith-CCM}. Further, when 
twinning occurs on limited twin systems, nanopillars undergo complete reorientation \cite{Rohith-CCM,Park-PRL}, which also 
leads to sequential reorientation\cite{Rohith-PML}, shape memory and pseudo-elasticity. However, when twinning occurs on 
multiple twin systems, it is possible that the twins belonging to different twin systems interact and form a more complex 
twin-twin junctions. Compared to individual twin boundaries (TBs), twin-twin junctions are complex in nature and have a 
significant and distinct role on the physical and mechanical properties of materials. For example, the twin-twin junctions 
retard grain boundary migration and/or grain growth\cite{Srolovitz}, causes detwinning \cite{Sainath-crystals}, hinders the 
growth of existing twins and increases the strain hardening rate \cite{El-Kadiri}. Such 
junctions can also influence the transport phenomena (ex. diffusion) along the grain boundaries \cite{Randle-Microscopy} and 
thus influencing the high temperature properties of materials like creep. Further, twin boundaries forming a twin-twin junction 
can accommodate large shear strains and lead to rearrangement of grain boundary network \cite{ACS-Nano,Wang-Ultramicroscopy}. 
Therefore, characterizing the twin junctions or twin-twin interactions is of utmost importance in understanding the micro-structural 
evolution and deformation mechanisms, which in turn dictates the mechanical properties.

There are many studies which characterized the twin-twin interactions in bulk materials \cite{Hadfield,Mullner}. For 
example, in Hadfield steel, it has been shown that, due to twin-twin interactions, the intersected region of two twins can 
form a second-order twin \cite{Hadfield}. A second-order twin is a twin formed within an already twinned region of a crystal. 
Similarly, many grain boundary engineering studies have shown that, when two coherent TBs interact, either $\Sigma$9 or 
$\Sigma$27 boundary is observed at their intersection, i.e., $\Sigma$3 + $\Sigma$3 = $\Sigma3^n$, where n = 1,2,3...
\cite{Randle-Microscopy,Randle-Acta2006}. It shows that the $\Sigma3^n$ type grain boundaries are geometrically related 
to $\Sigma$3 boundaries. Twin-twin interactions/junctions have also been characterized in nanocrystalline materials 
\cite{Srolovitz,Wang-Ultramicroscopy,ZH-Cao-Acta15}. Different twin junctions containing five, four, three and two 
TBs along with other grain boundaries have been observed. These junctions were characterized using the molecular dynamics 
simulations \cite{Srolovitz} and also using high resolution transmission microscopy (HRTEM) \cite{Wang-Ultramicroscopy}. 
Twin junctions with more than five TBs are not feasible due to crystallographic restriction. Out of all these junctions, 
observation of five-fold twin has attracted significant attention not only in plastic deformation of materials but also in 
crystallography and crystal growth studies. 

Like bulk and nanocrystalline materials, it is also important to understand the twin-twin interactions and junctions 
in nanopillars/nanoparticles, which show extensive twinning during deformation. However, surprisingly, there are no 
experimental reports of twin-twin interactions/junctions in nanopillars and also only a few simulation studies exist in 
characterizing the twin-twin junctions \cite{Conjoint,Torsion-5T}. These studies on nanopillars have shown only the formation 
of five-fold twins under the bending and torsional loading conditions \cite{Conjoint,Torsion-5T}. Further, most of 
the studies pertaining to the twin-twin interactions in bulk/nanocrystalline materials were experimental 
\cite{Wang-Ultramicroscopy,Hadfield,Mullner,ZH-Cao-Acta15}, where it is difficult to obtain the atomistic details. 
In view of this, the aim of the present investigation is to characterize and provide the atomistic mechanisms of how twin-twin 
interactions can lead to different twin-twin junctions in Cu nanopillars. We report two different twin-twin interactions, 
one containing two TBs along with one $\Sigma$9 boundary and the other with five TBs (five-fold twin), which are observed 
during the deformation of Cu nanopillars. The detailed dislocation/atomistic mechanisms responsible for the formation of 
such junctions have been explained by taking the advantage of atomistic simulations. 
  
\section{Simulation Details}

Single crystal Cu nanopillars oriented in [100] and $[1\bar1 0]$ axial directions were considered in this study. [100] nanopillar 
is enclosed by all \{100\} type side surfaces, while the $[1\bar10]$ nanopillar has (111) and $(11\bar2)$ as side surfaces. The 
model nanopillars had a square cross-section width (d) varying in the range 5.0 - 21.5 nm. The pillar length was twice the 
cross-section width. No periodic boundary conditions were used in any direction. On these model nanopillars, tensile loading 
has been simulated using molecular dynamics (MD) simulations. All MD simulations were carried out in LAMMPS package 
\cite{Plimpton-1995} employing an EAM potential for Cu given by Mishin et al. \cite{Mishin-2001}. This potential is widely 
used to study the deformation behaviour of Cu nanopillars \cite{Upmanyu,Sainath-PLA}. After initial construction of 
nanopillars, energy minimization was performed by conjugate gradient method to obtain a relaxed structure. Velocity verlet 
algorithm was used to integrate the equations of motion with a time step of 2 fs. Before applying tensile load, the relaxed 
nanopillars were equilibrated to a required temperature of 10 K in NVT ensemble. Following equilibration, the deformation was 
carried out in a displacement-controlled manner at constant strain rate of $1\times10^8$ s$^{-1}$ by imposing displacements 
to atoms along the nanopillar length that varied linearly from zero at the bottom fixed layer to a maximum value at the top 
fixed layer\cite{Sainath-PLA}. The visualization of TBs and dislocations is accomplished in AtomEye \cite{AtomEye} and 
OVITO \cite{OVITO} using common neighbour analysis (CNA). \hl{Further, all over the manuscript the TB means $\Sigma$3(111) 
coherent twin boundary}. 

\section{Results and Discussion}

\subsection{Formation of $\Sigma$9 boundary}

\begin{figure}
\centering
\includegraphics[width=11cm]{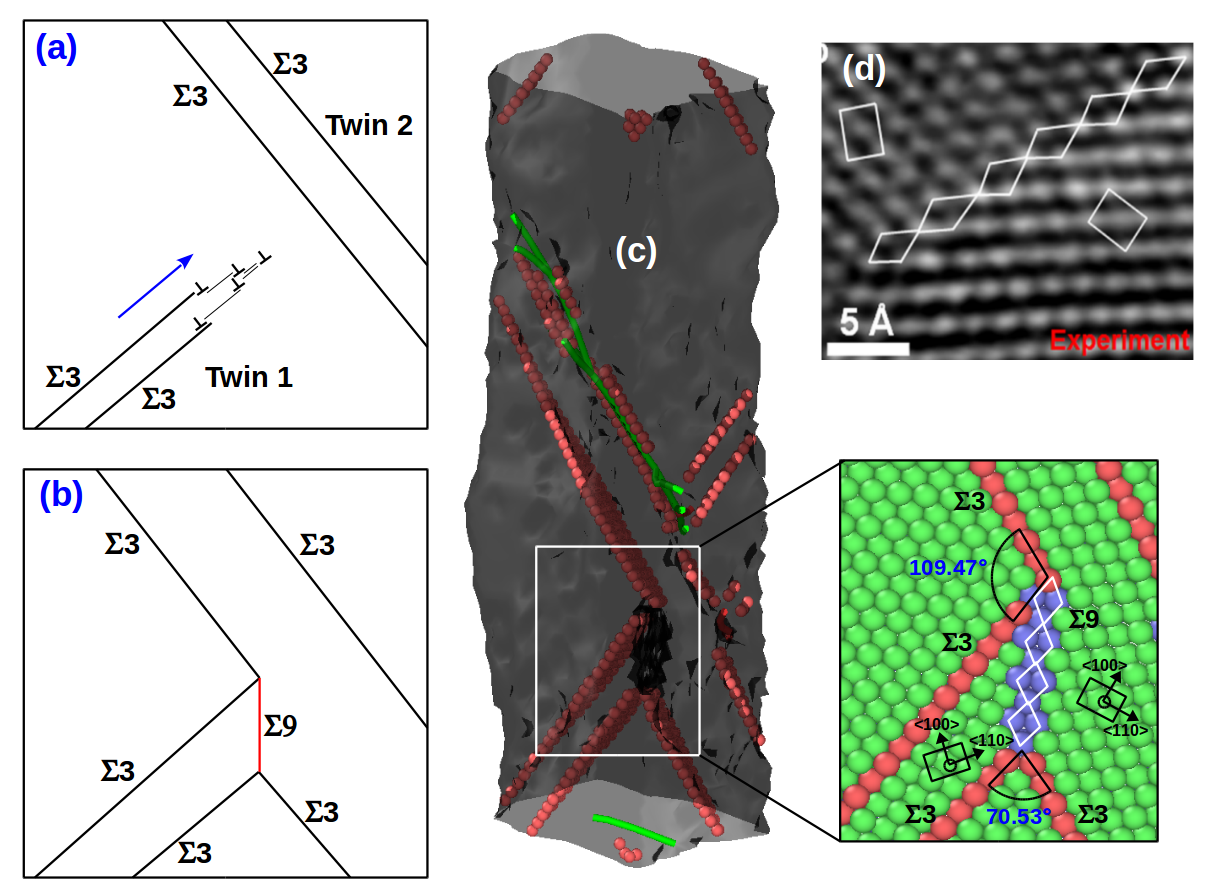}
\caption { (a-b) A schematic showing twin-twin interactions resulting in the formation of a $\Sigma$9 boundary, (c) formation of a 
$\Sigma$9 boundary at twin-twin junction during the tensile deformation of [100] Cu nanopillar with d = 10 nm, and (d) the structure 
of $\Sigma$9 boundary in nanocrystalline Pt observed using HRTEM by Wang et al. \cite{Wang-Ultramicroscopy}. The atoms in Figure 
(c) and its subset are colored according to their CNA parameter; \hl{green = FCC, orange = HCP (TBs) and blue = $\Sigma$9 boundary. 
The continuum green lines are partial dislocations.}}
\label{Fig01}
\end{figure}

Figure \ref{Fig01}a shows a schematic of twin (Twin-1) approaching towards an already existing twin (Twin-2) in FCC material. 
As a result of their interaction, a $\Sigma$9 boundary forms at their intersection, which connects the two $\Sigma$3-$\Sigma$3 
junctions (Figure \ref{Fig01}b). At one junction, the two $\Sigma$3 boundaries are at an acute angle ($70.53^o$) to each other, 
while at other junction, they make an obtuse angle ($109.47^o$). Interestingly, as shown in Figure \ref{Fig01}c, a similar 
twin-twin interaction leading to the formation of a $\Sigma$9 boundary has been observed under the tensile loading of $<$100$>$ 
Cu nanopillar. The detailed atomic structure of the observed $\Sigma$9 boundary viewed along $<$110$>$ misorientation axis is 
shown in subset Figure \ref{Fig01}c. It is interesting to see that the atomic structure of $\Sigma$9 boundary observed in the 
present investigation is quite similar to that observed using HRTEM (Figure \ref{Fig01}d) in nanocrystalline Pt \cite{Wang-Ultramicroscopy}. 
The misorientation angle (angle between \{110\} or \{100\} planes across the grain boundary) with respect to $<$110$>$ misorientation 
axis is found to be close to $42^o$, which is little higher than the theoretical value of $38.94^o$ \cite{PRB-GB}. This difference 
in misorientation angle can be attributed to local stress concentration and also deformation induced lattice distortion 
\cite{Wang-Ultramicroscopy}, which are neglected while calculating the theoretical values. Further, it has been found that the 
$\Sigma$9 boundary in Cu nanopillar (Figure \ref{Fig01}c) 
is parallel to one of the \{221\} planes. This abides well with the fact that the $\Sigma$9 boundary lies parallel to either 
\{114\} or \{122\} plane \cite{PRB-GB}.  

It is well known that when two $\Sigma$3 boundaries interact, either $\Sigma$9 or $\Sigma$27 boundary is observed at their 
intersection\cite{Randle-Acta2006}. However, the detailed mechanism responsible for the formation of such higher order boundary is 
not well understood. Understanding this mechanism is important as it is known that the $\Sigma$9 boundary takes part in reconfiguration 
of the existing grain boundary network during the plastic deformation.

\begin{figure}
\centering
\includegraphics[width=10cm]{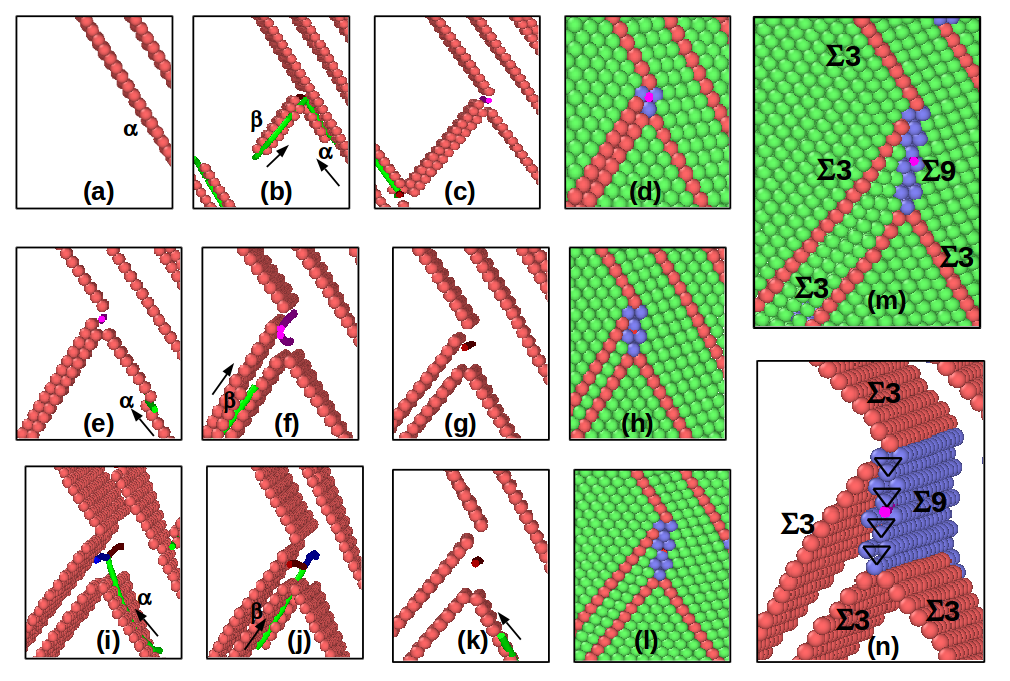}
\caption {(a-l) A detailed dislocation mechanism responsible for the formation of a $\Sigma$9 boundary during the tensile 
deformation of [100] Cu nanopillar, and (m-n) $\Sigma$9 boundary with four structural units connecting two twin-twin junctions. 
All the atoms are colored according to their CNA parameter; \hl{green = FCC, orange = HCP (TBs) and blue = $\Sigma$9 boundary. 
The continuum green lines are partial dislocations and other color lines are stair-rod or grain boundary dislocations.}}
\label{Fig02}
\end{figure}

Figure \ref{Fig02} shows the detailed mechanism responsible for the formation of a $\Sigma$9 boundary at the intersection of two 
$\Sigma$3 boundaries. As shown in Figure \ref{Fig02}a, an initial twin exists on plane $\alpha$ = (111). During 
deformation, a new slip system on plane $\beta$ = $(\bar{1}11)$ gets activated by the glide of a partial dislocation (Figure 
\ref{Fig02}b). With 
increasing strain, the partial dislocations on plane $\alpha$ and $\beta$ interact at the junction and forms a $\frac{1}{6}[011]$ 
stair-rod dislocation as shown in Figure \ref{Fig02}c-d. This reaction can be written as 
\begin{equation}
\frac{1}{6}[12\bar{1}]\beta + \frac{1}{6}[\bar{1}\bar{1}2]\alpha \longrightarrow \frac{1}{6}[011].
\end{equation}As shown in Figure \ref{Fig02}d, this stair-rod dislocation is surrounded by three atomic rows (blue colour), which 
are neither FCC atoms nor stacking fault (HCP) atoms. These three atomic rows form one structural unit of a grain boundary. 
Following the formation of one structural unit, a twinning partial dislocation glides on plane $\alpha$ (Figure \ref{Fig02}e) 
and interacts with $\frac{1}{6}[011]$ stair-rod dislocation (Figure \ref{Fig02}f) according to the reaction 
\begin{equation}
 \frac{1}{6}[\bar{1}\bar{1}2]\alpha + \frac{1}{6}[011] \longrightarrow \frac{1}{6}[\bar{1}03].
\end{equation}Here, the reaction violates the Frank rule which makes the $\frac{1}{6}[\bar{1}03]$ dislocation highly unstable. 
As a result, this dislocation frequently dissociates (Figure \ref{Fig02}f) and recombines according to the following reaction
\begin{equation}
 \frac{1}{6}[\bar{1}03] \longrightarrow \frac{1}{6}[0\bar{1}\bar{1}] + \frac{1}{6}[\bar{1}14] \longrightarrow \frac{1}{6}[\bar{1}03].
\end{equation}Interestingly, the $\frac{1}{6}[\bar{1}03]$ dislocation formed via above recombination reaction minimises the energy, 
i.e., follows the Frank rule and also it is stable, unlike the previous one. Following recombination, another partial dislocation 
gliding on plane $\beta$ interacts with $\frac{1}{6}[\bar{1}03]$ dislocation and results in the formation of another new stair-rod 
dislocation as shown in Figure \ref{Fig02}g-h. This reaction can be written as \begin{equation}
 \frac{1}{6}[\bar{1}03] + \frac{1}{6}[12\bar{1}]\beta \longrightarrow \frac{1}{6}[022] \longrightarrow \frac{1}{3}[011].
\end{equation}As shown in Figure \ref{Fig02}h, the reactions (2) and (4) together adds one more structural unit to the grain 
boundary. The $\frac{1}{3}[011]$ stair-rod dislocations formed at the end of reaction (4) lies in between these two structural 
units (Figure \ref{Fig02}h) and, it is equivalent of two $\frac{1}{6}[011]$ stair-rod dislocations ($\frac{1}{3}[011] \longrightarrow 
\frac{1}{6}[011] + \frac{1}{6}[011]$), each at one structural unit. With further deformation, one more partial dislocation gliding 
on plane $\alpha$ comes and interacts with $\frac{1}{3}[011]$ stair-rod dislocation (Figure \ref{Fig02}i) and results in the formation 
of a new dislocation as follows 
\begin{equation}
 \frac{1}{6}[\bar{1}\bar{1}2]\alpha + \frac{1}{6}[022] \longrightarrow \frac{1}{6}[\bar{1}14]
\end{equation}Here, the resultant $\frac{1}{6}[\bar{1}14]$ dislocation forms a multiple of $\frac{1}{18}[\bar{1}14]$, which is one 
of the displacement shift complete (DSC) lattice dislocation of a $\Sigma$9 boundary \cite{Foll-defects}. Also, the reaction 
increases the energy, i.e., violates Frank criterion. However, the reaction is feasible due to the presence of high stresses within 
the grain boundary. Subsequently, one more partial dislocation gliding on plane $\beta$ also comes and interacts with $\frac{1}{6}[\bar{1}14]$ 
dislocation (Figure \ref{Fig02}j) and results in the formation of $\frac{1}{2}[011]$ stair-rod dislocation (Figure \ref{Fig02}k-l) 
according to the following reaction 
\begin{equation}
 \frac{1}{6}[12\bar{1}]\beta + \frac{1}{6}[\bar{1}14]  \longrightarrow \frac{1}{2}[011]
\end{equation}The reactions (5) and (6) together add one more structural unit to the grain boundary as shown in Figure 
\ref{Fig02}l. Here, the $\frac{1}{2}[011]$ stair-rod dislocation is equivalent of three $\frac{1}{6}[011]$ stair-rod dislocations 
($\frac{1}{2}[011]  \longrightarrow  \frac{1}{6}[011] + \frac{1}{6}[011] + \frac{1}{6}[011]$), one at each structural unit. Thus, 
every interaction of partial dislocations emanating from two intersecting \{111\} planes contributes one $\frac{1}{6}[011]$ stair-rod 
dislocation and adds one structural unit. In other words, the formation of $\Sigma$9 boundary just needs the repeated generation 
of 1/6$<$112$>$ dislocation locks though two partial dislocation reactions as shown in Figures \ref{Fig02}a-d. The continuous 
repetition of this process adds more and more structural units leading to an increase in $\Sigma$9 grain boundary length 
(Figure \ref{Fig02}m-n). These results provide the direct evidence to the fact that $\Sigma$9 boundary is an array of 
$\frac{1}{6}[011]$ type stair-rod dislocations as suggested by Zhu and co-workers \cite{Zhu-Twin-twin}.

\subsection{Formation of five-fold twin}

\begin{figure}
\centering
\includegraphics[width=7.5cm]{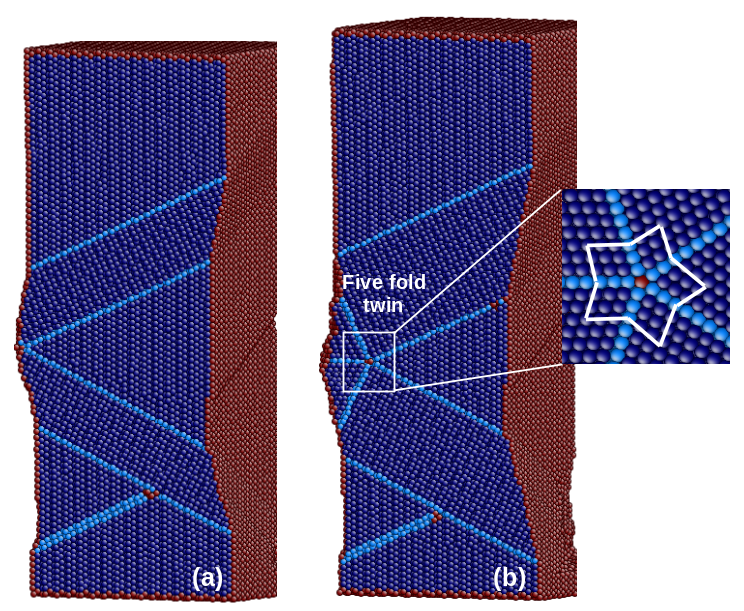}
\caption {Twin-twin interactions leading to the formation of five-fold twin during the tensile deformation of $<$110$>$ Cu 
nanopillar with d = 7.2 nm. (a) Two twins interacting at the nanopillar corner and (b) five-fold twin formed at the intersection 
of two twin boundaries near nanopillar corner. The atoms are colored according to their CNA parameter; \hl{dark blue = FCC, light 
blue = HCP (TBs) and red = surface or dislocation core atoms.}}
\label{Fig03}
\end{figure}

Five-fold twin is basically a junction of five different TBs meeting at one point and it exhibits an interesting 
five-fold symmetry. Twin-twin interactions leading to the formation of five-fold twin during the tensile deformation of $<$110$>$ 
Cu nanopillar is shown in Figure \ref{Fig03}. Initially, the two twins (twin-1 and twin-2) grow on two different planes (Figure 
\ref{Fig03}a) and as result of their interaction at the nanopillar corner, a five-fold twin forms at the intersection of two 
leading TBs (Figure \ref{Fig03}b). Similar to the present observation, few studies in the literature have reported the formation 
of five-fold twins in nanopillars under bending and torsional loading conditions \cite{Conjoint,Torsion-5T}. However, 
it has never been observed during the uniaxial loading of nanopillars. Initially, it has been proposed that the essential 
requirement for the formation of five-fold twin are (i) large shear stress to activate multiple slip systems and (ii) variation 
in stress orientation such that several partials with different orientations can be nucleated \cite{Zhu-APL,Bringa-Scripta}. 
Generally, torsion and bending tests provide such conditions. Accordingly, it was claimed that the uniaxial stress conditions 
may not lead to five-fold twins \cite{Zhu-APL}. However, the observation of five-fold twin close to nanopillar corner as shown 
in Figure \ref{Fig03}b indicates that these two conditions were readily satisfied at nanopillar corners even under tensile 
loading. Generally, the stress is high at nanopillars corners and also there is a possibility of change in stress orientation. 
Similar to the present study, Cao and Wei \cite{Cao-Wei-APL} have also reported the formation of five-fold twins under tensile 
loading of nanocrystalline Cu and suggested that the stress state within a small region is always a complex one even under 
uniaxial conditions. 

In literature, different mechanisms have been suggested for the formation of five-fold twin. In nanocrystalline materials, 
it has been shown that the migration of grain boundary segment is responsible for the formation of five-fold twin \cite{Srolovitz}. 
Similarly, Bringa et al.\cite{Bringa-Scripta} have reported that multiple emissions of partial dislocations from the grain 
boundaries due to high local stresses leads to the formation of five-fold twin. In nanopillars, Zheng et al. \cite{Conjoint} 
have demonstrated that an intermediate icosahedral phase facilitates the formation of five-fold twin. In Cu nanoparticles, the 
TEM observations indicate that initially the lattice gets distorted and this distorted portion acts as a partial dislocation 
source \cite{FT-Cu-NPs}. The glide of these partial dislocations induces a layer-by-layer migration of TBs resulting in a standard 
five-fold twin \cite{FT-Cu-NPs}. Thus, there lies a great ambiguity over the formation of five-fold twin in nanopillars/nanocrystalline 
materials.

\begin{figure}
\centering
\includegraphics[width=9cm]{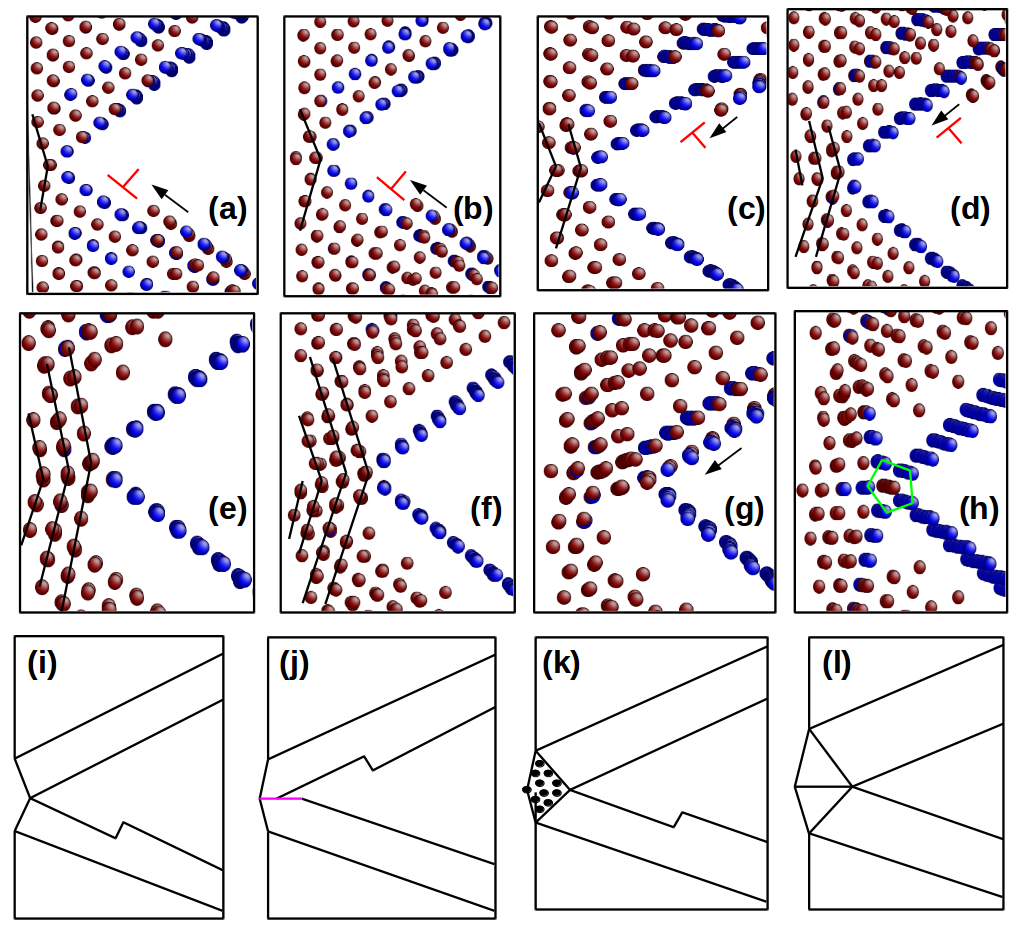}
\caption {(a-h) The detailed atomistic mechanism responsible for the formation of five-fold twin during the tensile deformation 
of $<$110$>$ Cu nanopillar. The same process is summarized schematically in (i-l). The atoms in Figure (a-h) are colored according 
to their CNA parameter; \hl{blue = HCP (TBs) and red = surface or dislocation core atoms. The FCC atoms are not shown for clarity.}}
\label{Fig04}
\end{figure}

The detailed atomistic mechanism responsible for the formation of five-fold twin in the present study is shown in Figure \ref{Fig04}. 
Different from previous observations, five-fold twin directly nucleates from the distorted lattice formed at nanopillar corner. During 
initial straining, two TBs making an angle of $70.53^o$ with respect to each other meet at nanopillar corner 
(Figure \ref{Fig03}a and \ref{Fig04}a). Following this, a partial dislocation glides on one of the TBs, which drives the twin junction 
by one atomic step away from corner (Figure \ref{Fig04}a-b). This process of partial dislocation glide continues to occur on two 
intersecting TBs and makes the twin junction to move inwards, i.e., away from nanopillar corner (Figure \ref{Fig04}a-d). 
As a result, the atoms/planes between nanopillar corner and twin junction change to disorder/non-closed pack state (Figure \ref{Fig04}c-f). 
In other words, the lattice gets distorted locally near nanopillar corner due to twin-twin interactions. With increasing strain, as 
this distorted lattice is highly unstable, a five-fold twin nucleates from the distorted region due to atomic readjustments (Figure 
\ref{Fig04}g-h). These atomic adjustments were mainly aided by the partial dislocation movements on the existing TBs. 
As a result, a clear five-fold twin can be seen in Figure \ref{Fig04}h. This complete process is summarized schematically in Figures 
\ref{Fig04}i-l. Since the angle between any \{111\} planes in FCC system is $70.53^o$, a five-fold twin leaves a gap of $7.35^o$ 
$(360 -5\times70.53)$. However, such gap is accommodated by the elastic strain within the twinned regions. It has been further 
observed that, with increasing deformation, the centre point of five-fold twin move towards the nanopillar interior due to partial 
dislocation glide on TBs. Thus, the five-fold twin under the tensile loading of Cu nanopillars nucleates from the distorted lattice 
formed close to nanopillar corner. The distorted lattice at the corner is formed due to the continuous glide of partial dislocations 
on two intersecting TBs. This process of five-fold twin formation is remarkably different from that reported in nanocrystalline 
materials \cite{Srolovitz,Bringa-Scripta} and other nanopillars/nanoparticles \cite{Conjoint,FT-Cu-NPs}. In previous studies, the 
involvement of grain boundaries/partial dislocations/icosahedral phase was needed for five-fold twin formation, whereas the present 
study shows that the atomic shuffling at the intersection point of two existing twin boundaries can also results 
in five-fold twin formation.

\subsection{Size effects and stress-strain behaviour}
The above mentioned twin-twin junctions, i.e., $\Sigma$3-$\Sigma$3-$\Sigma$9 and five-fold twin, have been observed in all the 
nanopillars with size (d) in the range 5.0 - 21.5 nm. However, the area (or structural units in \hl{Figure \mbox{\ref{Fig02}}n)} 
of the newly formed $\Sigma$9 boundary is different in different nanopillars. The observed number of structural units as shown 
in \hl{Figure \mbox{\ref{Fig02}}n} varied from two to maximum of six. On the contrary, the five-fold twin has not shown any remarkable 
differences among nanopillars of different size. However, it is of significant interest to understand the size dependence of 
twin-twin junctions in nanopillars and nanocrystalline materials over a wide range of sizes. To this, Cao et al. 
\cite{ZH-Cao-Acta15} have analysed the formation frequency of different twin-twin junctions with respect to the grain 
size in nanocrystalline Cu. It has been found that formation frequency increases with decreasing grain size, then reaches a peak 
around 35-45 nm, and again decreases at lower grain sizes \cite{ZH-Cao-Acta15}. Similar study is needed in nanopillars 
to understand the size dependence of twin-twin junctions.

\begin{figure}[h]
\centering
\includegraphics[width=9cm]{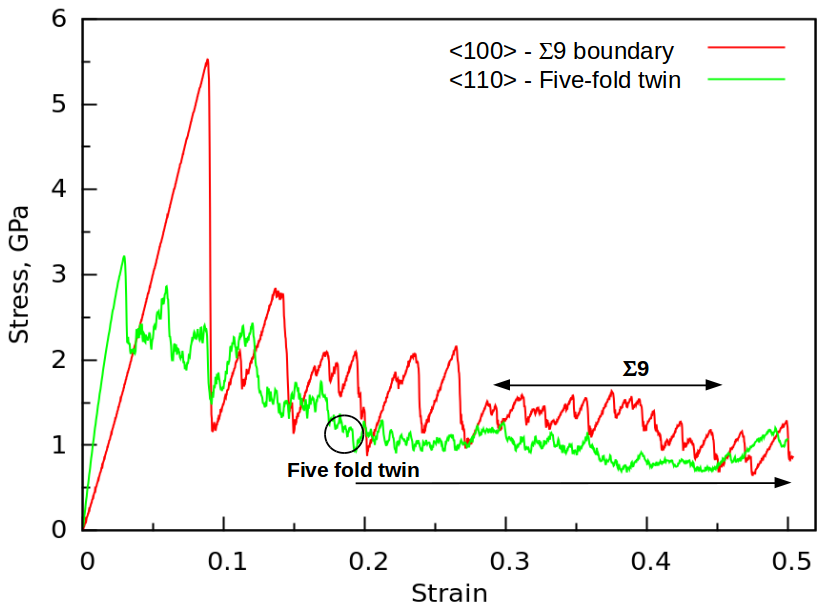}
\caption {The stress-strain behaviour of [100] and $[1\bar1 0]$ Cu nanopillars under tensile loading at 10 K. In [100] nanopillar, 
the formation of $\Sigma$9 boundary has been observed, while in $[1\bar1 0]$ nanopillar, a five-fold twin was observed. The strain 
values at which the twin-twin interactions were observed are indicated.}
\label{Fig05}
\end{figure}

Figure \ref{Fig05} shows the stress-strain behaviour of [100] and $[1\bar1 0]$ Cu nanopillars with d = 10 nm, where the 
formation of $\Sigma$9 boundary and five-fold twin was observed in respective orientations. Following an initial linear elastic 
deformation, both the nanopillars show an extensive plastic deformation characterized by flow stress fluctuations along with an 
average decrease in flow stress. The strain range over which the above mentioned twin-twin interactions were observed is 
highlighted in Figure \ref{Fig05}. The formation of $\Sigma$9 boundary has been observed over a strain range of 0.3-0.45. 
Over this strain range, the area of the grain boundary increases gradually with increasing strain beginning with one structural 
unit. On the other hand, the five-fold twin boundary has formed at a strain ($\varepsilon)$ value of 0.18 and remained in the 
nanopillar till the strain value of more than 0.5. During this period ($\varepsilon = 0.18-0.5$), the gradual movement of centre 
of five-fold twin has been observed.

\section{Conclusions}

In summary, molecular dynamics simulations performed on [100] and $[1\bar1 0]$ Cu nanopillars indicate that multiple twin systems 
activate and interactions among themselves leads to complex twin-twin junctions. The atomistic mechanisms responsible for the 
formation of such twin-twin junctions have been revealed in detail. During the tensile deformation of [100] nanopillar, a twin junction 
containing two TBs along with one $\Sigma$9 boundary, i.e.,$\Sigma$3-$\Sigma$3-$\Sigma$9 junction, has been observed. At this junction, 
the $\Sigma$9 boundary is added unit by unit by an interesting partial dislocation reaction at the intersection of two TBs. This process 
is quite similar to the experimental observation in nanocrystalline Pt \cite{Wang-Ultramicroscopy}. However, due to the advantage of 
atomistic simulations, the present study provides the detailed picture of dislocation interactions 
at the twin-twin junction. This understanding help us to explain how the $\Sigma$9 boundary takes part in reconfiguring the grain 
boundary network during the plastic deformation of polycrystalline materials. On the other hand, in $[1\bar1 0]$ nanopillar, a 
junction of five different TBs meeting at one point, i.e., five-fold twin has been observed. This five-fold twin nucleates from the 
distorted lattice formed close to nanopillar corner. The distorted lattice at the corner is formed due to the continuous glide of 
partial dislocations on the two intersecting TBs. This observation sheds a new light on the formation mechanism of five-fold 
twin in nanopillars. 



\end{document}